\documentclass[journal=jacsat,manuscript=article,layout=twocolumn]{achemso} 
\usepackage{epsfig}
\usepackage{color}
\usepackage{amsmath}
\usepackage{amssymb}
\usepackage{amsfonts}
\usepackage{graphicx}

\usepackage[T1]{fontenc}       
\usepackage{siunitx}
\usepackage{verbatim}

\newcommand{\kbt}{ k_{\rm B}T }
\newcommand{\Nvphi}{ \langle N_v \rangle_{\phi} }
\newcommand{\suscep}{ -\partial \langle N_v \rangle_{\phi} / \partial (\beta\phi)}
\newcommand{\cv}{ \chi_v }

\newcommand{\supp}{\mathrm{Supporting~Information}}

\author{Nicholas B. Rego}
\affiliation{Biochemistry \& Molecular Biophysics Graduate Group, University of Pennsylvania, Philadelphia, PA 19104}

\author{Erte Xi}
\affiliation{Department of Chemical \& Biomolecular Engineering, University of Pennsylvania, Philadelphia, PA 19104}

\author{Amish J. Patel}
\affiliation{Department of Chemical \& Biomolecular Engineering, University of Pennsylvania, Philadelphia, PA 19104}
\email{amish.patel@seas.upenn.edu}

\title{Protein Hydration Waters are Susceptible to Unfavorable Perturbations}


\begin{document} 

%
\begin{tocentry}
\includegraphics[height=1.75in]{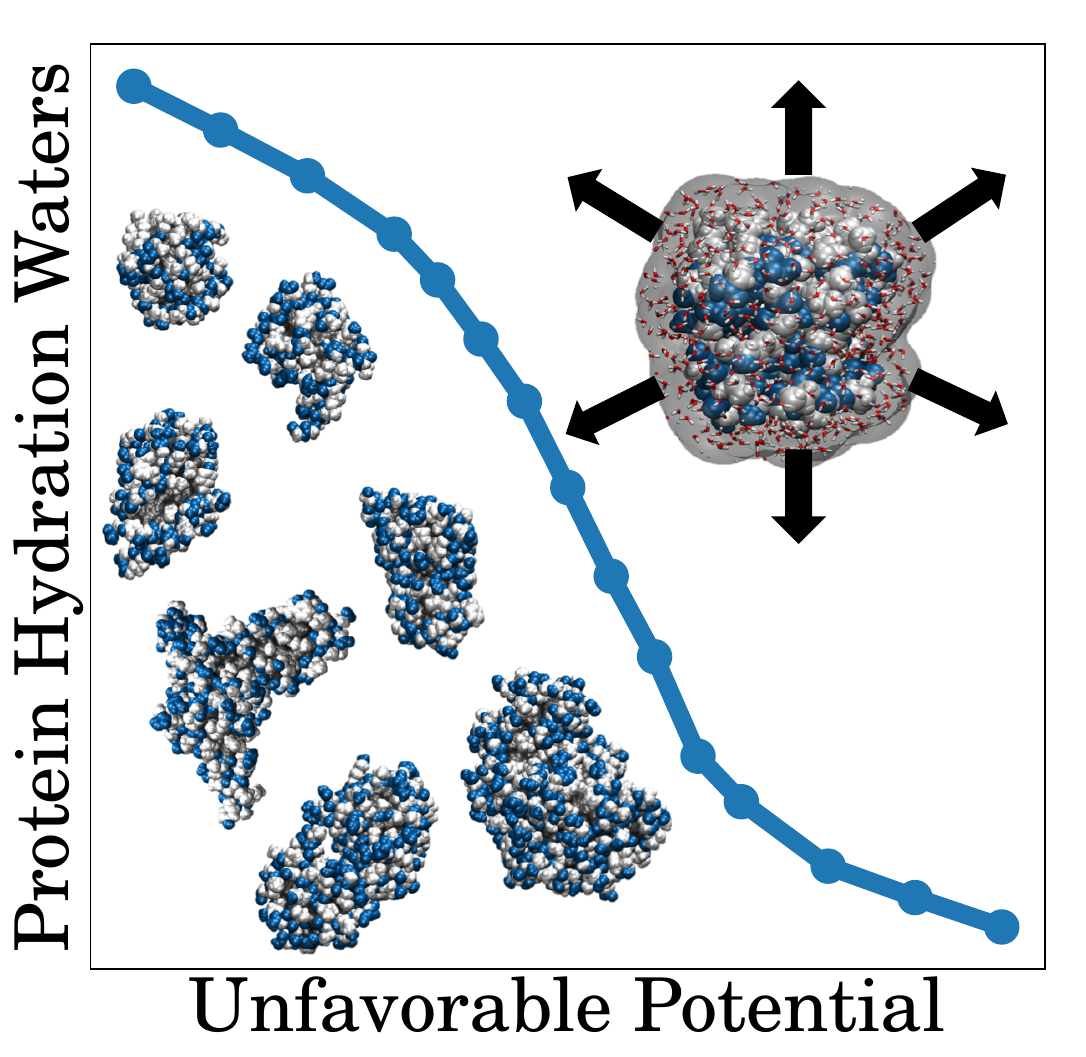} 
\end{tocentry}

\maketitle

\begin{abstract}
The interactions of a protein, its phase behavior, and ultimately, its ability to function, 
are all influenced by the interactions between the protein and its hydration waters.
Here we study proteins with a variety of sizes, shapes, chemistries, and biological functions, 
and characterize their interactions with their hydration waters 
using molecular simulation and enhanced sampling techniques. 
We find that akin to extended hydrophobic surfaces,
proteins situate their hydration waters at the edge of a dewetting transition,
making them susceptible to unfavorable perturbations.
We also find that the strength of the unfavorable potential needed to trigger dewetting is roughly the same, 
regardless of the protein being studied, and depends only the width of the hydration shell being perturbed.
Our findings establish a framework for systematically classifying protein patches according to how favorably they interact with water.
\end{abstract}

\raggedbottom

\section{Introduction}
%
Biomolecular binding processes involve disrupting protein-water interactions 
and replacing them with direct interactions between the binding partners. 
Thus, protein-water interactions influence the thermodynamics and kinetics of protein interactions~\cite{Ashbaugh:2008:JACS,shea08,levy2006water,berne_rev09,Jamadagni:ARCB:2011,Abrams:2013:PSFB,Baron:2013aa,tiwary2015kinetics,tiwary2015role,bellissent2016water},
as well as the stability and phase behavior of protein solutions~\cite{Jaenicke:2000hb,Shire:JPS:2004,Shen:2006aa,Trevino:2008jh,Thirumalai:2012,Palmer:JPCL:2012,Ham:Angw:2014,Remsing:JPCB:2018}.
In this article, we characterize the 
overall interactions between proteins and their hydration waters 
by using specialized molecular simulations that employ an unfavorable potential 
to displace water molecules from the vicinity of a protein.
Because displacing interfacial waters disrupts surface-water interactions, 
the less favorable those interactions (e.g., for hydrophobic surfaces),
the easier it is to displace the interfacial waters~\cite{Patel:PNAS:2011,Patel:JPCB:2014,Xi:JCTC:2016}.
Indeed, both theory~\cite{LCW,LLCW,vaikuntanathan2016necessity,Xi:PNAS:2016}
and molecular simulations~\cite{Patel:JPCB:2010,Patel:JPCB:2012} have shown that the rare, low-density fluctuations,
which are accessed when interfacial waters are displaced,
are substantially more probable adjacent to a hydrophobic surface than at a hydrophilic surface.
Moreover, water molecules near a hydrophobic surface are susceptible to unfavorable perturbations, 
and undergo a collective dewetting transition in response to such a perturbation~\cite{LCW,Patel:JPCB:2010,Patel:JPCB:2012,Xi:JCTC:2016}.
%
Proximity to such a dewetting transition is also reflected in other collective interfacial properties, such as
compressibility, transverse density correlations, and the distribution of water dipole orientations, among others
~\cite{LeeMR,LCW,mittal_pnas08,Sarupria:PRL:2009,Godawat:PNAS:2009,Patel:JPCB:2010,Willard:JPCB:2010,Patel:PNAS:2011,Patel:JPCB:2012,heyden2013spatial,sosso2016role,Shin:JPCB:2018}.

In contrast with simple hydrophobic or hydrophilic surfaces, proteins display nanoscopic chemical and topographical patterns,
which influence their interactions with water in non-trivial ways~\cite{Giovambattista:PNAS:2008,giovambattista2009enhanced,Acharya:Faraday:2010,daub2010influence,mittal2010interfacial,Wang19042011,fogarty2014water,Patel:JPCB:2014,harris2014effects,Xi:PNAS:2017,Shell:PNAS:2018,heyden:2018}.
By interrogating how protein hydration waters respond to an unfavorable potential,
here we find that the hydration shells of diverse proteins are also situated at the edge of a dewetting transition. 
Such a resemblance of protein hydration shells to extended hydrophobic surfaces 
appears to arise from the fact that --
even for protein surfaces that are enriched in polar and charged residues --
roughly half the surface consists of hydrophobic atoms. 
Our findings, obtained by studying proteins across a broad range of sizes, chemistries, and functions,
suggest that susceptibility to unfavorable perturbations 
is a common feature of soluble proteins with well-defined folded structures.

We also find that the strength of the unfavorable potential needed to trigger dewetting is roughly the same across all proteins, 
and depends only on the width of the hydration shell  we choose to perturb.
Our findings lay the groundwork for systematically disrupting protein-water interactions, 
and uncovering regions of proteins that have the weakest (hydrophobic) and the strongest (hydrophilic) interactions with water.
A knowledge of the most hydrophobic protein regions 
could enable the prediction of the interfaces through which protein interact with one another~\cite{Bogan:JMB:1998,DeLano:2002aa,nooren2003diversity,White:2008aa}.
Similarly, uncovering the most hydrophilic protein patches could result in the discovery of novel super-hydrophilic chemical patterns~\cite{nonfouling}.

\begin{figure*}[htb]
\centering
\includegraphics[width=0.8\textwidth]{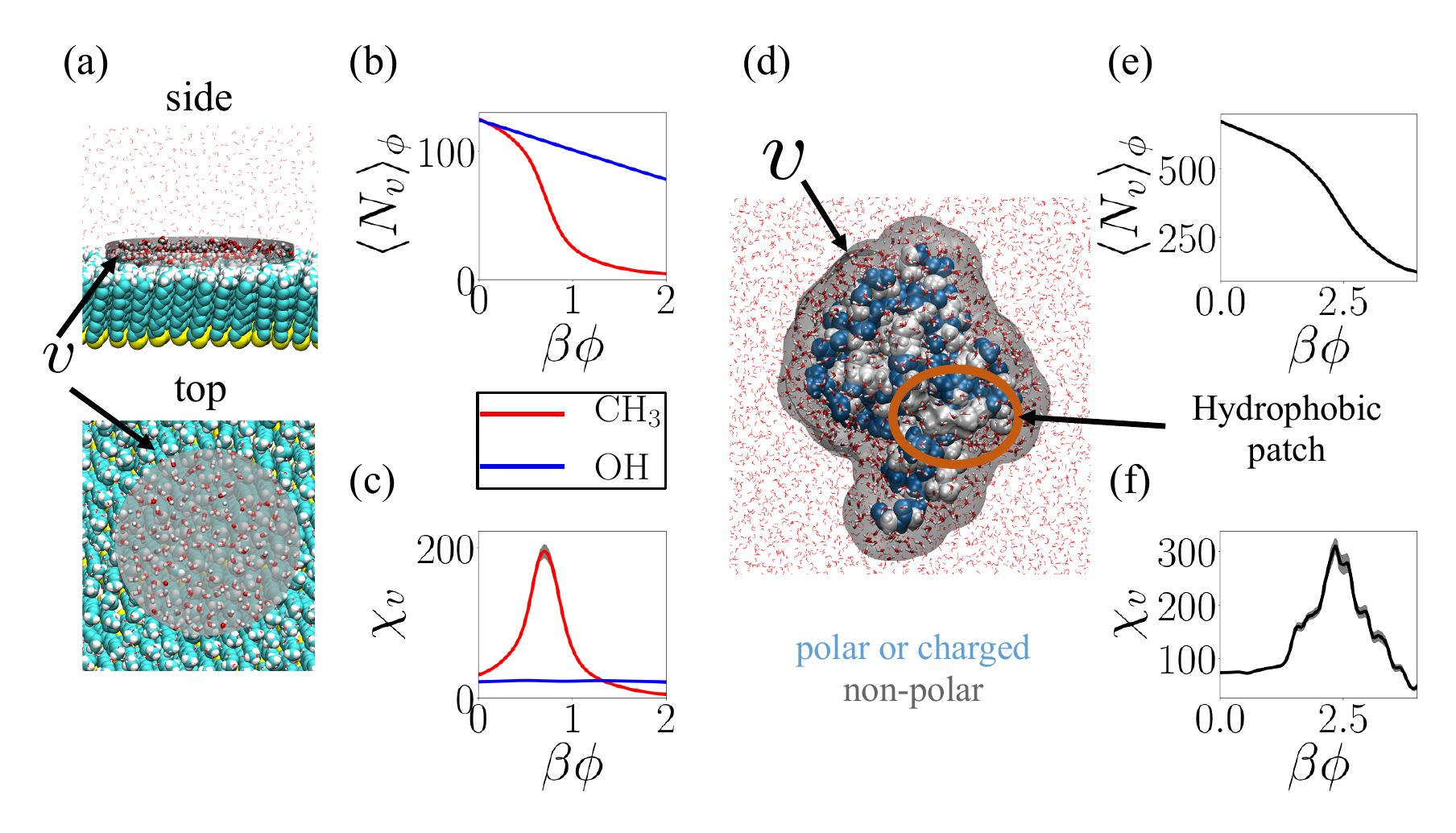} 
\vspace{-0.3in}
\caption{
How interfacial waters respond to an unfavorable potential.
(a) Simulation snapshots of the CH$_3$-terminated SAM-water interface.
The SAM atoms are shown in space-fill representation; 
the $N_v$ waters in the interfacial observation volume, $v$, are shown in the licorice representation, and the rest as lines.
The cylindrical $v$ is chosen to have a radius, $R_v = 2$~nm, and a width, $w = 0.3$~nm.
(b) In response to an unfavorable biasing potential, $\phi N_v$, the average number of interfacial waters, $\Nvphi$, 
decreases at both the hydrophobic CH$_3$-terminated and the hydrophilic OH-terminated SAM surfaces.
However, as the potential strength, $\phi$ is increased, $\Nvphi$ near the hydrophilic SAM decreases gradually;
whereas, $\Nvphi$ near the hydrophobic SAM decreases sharply~\cite{LCW,Patel:JPCB:2010,Patel:JPCB:2012,Xi:JCTC:2016}. 
(c) These differences are also evident in the $\phi$-dependence of the susceptibility, $\cv \equiv \suscep$,
which is roughly constant for the hydrophilic SAM, but shows a marked peak for the hydrophobic SAM.
Thus, interfacial waters at a hydrophobic surface are susceptible to unfavorable perturbations~\cite{LCW,LLCW,Patel:JPCB:2010,Patel:JPCB:2012}.
(d) Simulation snapshot of the ubiquitin protein, 
highlighting the observation volume, $v$, 
which contains waters in the first hydration shell of the protein.
Protein atoms are shown in space-fill representation, and colored according to their atom types (white = hydrophobic, blue = hydrophilic) following ref.~\cite{rossky};
waters in $v$ are shown in licorice representation, and the rest are shown as lines. 
The well-characterized hydrophobic patch of ubiquitin, which mediates its interactions with other proteins, 
is also shown~\cite{ubiq_recog}. 
(e) The average number of protein hydration waters, $\Nvphi$, decreases in a sigmoidal manner as $\phi$ is increased.
(f) The corresponding susceptibility, $\cv$, displays a maximum, suggesting that protein hydration waters 
are also susceptible to unfavorable perturbations, akin to those near hydrophobic surfaces. 
\vspace{-0.25in}
}
\label{fig1}
\end{figure*}

\section{Results and Discussion}
%
\subsection{Water near uniform, flat surfaces}
%
To illustrate the molecular signatures of surface hydrophobicity,
we first review the contrasting behavior of water near CH$_3$-terminated (hydrophobic) and OH-terminated (hydrophilic) 
self-assembled monolayer (SAM) surfaces.
The SAM surfaces are not only flat and uniform, and thereby considerably simpler than proteins,
but their hydrophobicity can also be defined unambiguously using a macroscopic measure, such as the water droplet contact angle.
We focus on water molecules in a cylindrical observation volume, $v$, at the SAM-water interface, as shown in Figure~\ref{fig1}a. 
We choose a radius, $R_v = 2$~nm, and a width, $w = 0.3$~nm, for the cylindrical $v$;
with this choice, $v$ at either SAM surface contains an average of roughly 120 waters. 
%
Following previous work~\cite{Patel:JSP:2011,Patel:JPCB:2012,Patel:JPCB:2014,Xi:JCTC:2016},
we then perturb the interfacial waters in $v$ by applying an unfavorable biasing potential, $\phi N_v$,
where $\phi$ represents the strength of the potential, 
and $N_v$ is the number of coarse-grained waters in $v$; 
a more precise definition of $N_v$ is included in the $\supp$.
The potential imposes an energetic penalty that increases linearly with $N_v$, 
so that
%
as $\phi$ is increased, waters are displaced from $v$, resulting in a decrease in the average water numbers, $\Nvphi$, next to both SAM surfaces;
see Figure~\ref{fig1}b. 
The decrease in $\Nvphi$ with increasing $\phi$ is linear for the hydrophilic SAM.
In comparison, the corresponding $\Nvphi$-values for the hydrophobic SAM are smaller for every $\phi$,
highlighting the relative ease of displacing waters.
Moreover, the decrease in $\Nvphi$ with increasing $\phi$ is sensitive (or sigmoidal) 
near the hydrophobic surface rather than gradual (and linear) as it is near the hydrophilic surface.
This contrast can be seen even more clearly in Figure~\ref{fig1}c, which shows the susceptibility, 
$\cv \equiv \suscep$, as a function of $\phi$; here, $\beta = 1 / \kbt$, $k_{\rm B}$ is the Boltzmann constant, and $T$ is the system temperature.
The susceptibility is nearly constant for the hydrophilic surface.
However, it shows a pronounced peak for the hydrophobic surface,
suggesting that a collective dewetting of the interfacial waters 
can be triggered when a sufficiently strong unfavorable potential is applied.
%

\subsection{Perturbing the protein hydration shell}
%
In contrast with the uniform SAM surfaces, proteins are heterogeneous, rugged, and amphiphilic.
Their surfaces tend to have polar and charged residues to ensure that the protein is soluble in water,
as well as hydrophobic residues to drive protein interactions.
Given the amphiphilicity of proteins, how might their hydration waters respond to an unfavorable potential?
Should we expect its hydration waters to be displaced gradually like the hydrophilic SAM surface?
Or should we expect the protein hydration waters to undergo collective dewetting like the hydrophobic SAM surface?
To address these questions, we first study ubiquitin, a highly-conserved protein involved in numerous signaling pathways, including protein degradation~\cite{ubiq_1}. 
Although it is a relatively small protein (76 amino acid residues), 
ubiquitin displays many of the characteristic features of a soluble globular protein, 
including a stable folded structure, a chemically and topographically heterogeneous surface, and 
interactions with diverse molecules that are crucial to its function (Figure~\ref{fig1}d)~\cite{ubiq_recog}.
Many of these interactions are mediated by a well-documented hydrophobic patch, 
which is also shown in Figure~\ref{fig1}d. 

To characterize the overall strength of protein-water interactions, we again apply a biasing potential, $\phi N_v$, 
where $N_v$ is now the number of coarse-grained waters in the entire protein hydration shell, $v$.
The hydration shell, $v$, is defined as the union of spherical sub-volumes centered on all the protein heavy atoms,
with each sub-volume chosen to have the same radius, $R_v$. 
Such a definition allows $v$ to capture the ruggedness of the underlying protein surface,
with the width of the hydration shell determined by $R_v$.
Here we choose $R_v=0.6$~nm so that only waters in the first hydration shell of the protein are included in $v$ (Figure~\ref{fig1}d).
The decrease in the average number of ubiquitin hydration waters, $\Nvphi$, in response to the unfavorable potential, $\phi N_v$, is shown in Figure~\ref{fig1}e.
Interestingly, $\Nvphi$ displays a sigmoidal dependence on $\phi$, akin to that for the hydrophobic CH$_3$-terminated SAM surface (Figure~\ref{fig1}b).
Correspondingly, a clear peak is also observed in the susceptibility around $\phi^* \approx 2~\kbt$ (Figure~\ref{fig1}f).
Thus, the hydration shell of the inherently amphiphilic and incredibly complex surface of the ubiquitin protein dewets collectively in response to an unfavorable perturbation. 

\begin{figure*}[thb]
\centering
\includegraphics[width=1.\textwidth]{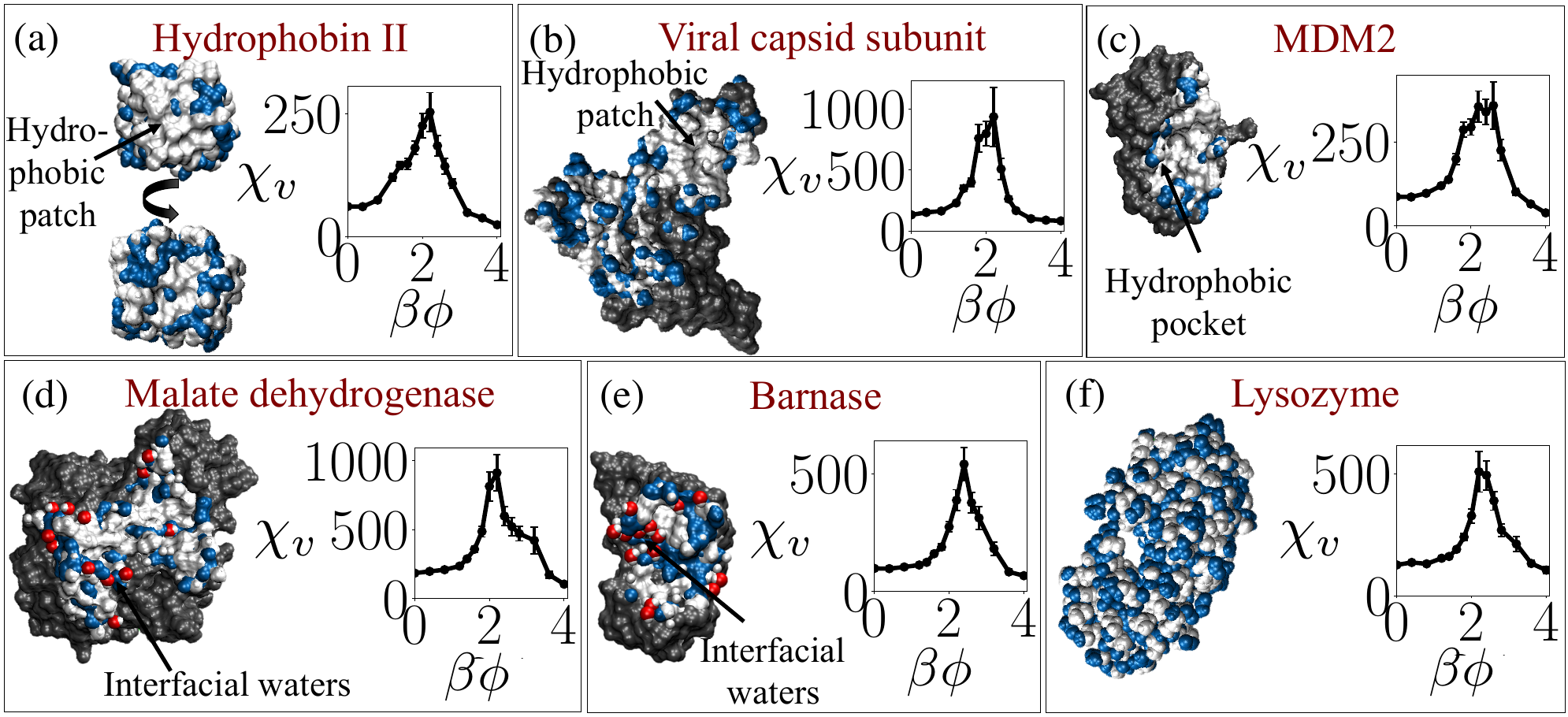} 
\vspace{-0.25in}
\caption{
The hydration shells of proteins with diverse sizes, shapes, chemical patterns, and functional roles are susceptible to an unfavorable potential.
For each protein, surface atoms of interest are shown in white (non-polar) or blue (polar or charged); 
the rest are shown in gray.
Susceptibilities of the protein hydration waters, $\cv \equiv \suscep$, to the biasing potential strength, $\phi$, 
are also shown. 
(a) Hydrophobin II is a small fungal protein that adsorbs to water-vapor interfaces via a large hydrophobic patch (top); 
the rest of the protein is amphiphilic (bottom)~\cite{hfb_struct,hfb_rev}.
(b)  The protein that makes up a structural sub-unit of the hepatitis B viral capsid is shown, 
along with the amphiphilic interface through which two protein sub-units bind; 
the hydrophobic patch on the binding interface is also highlighted.
(c) The signaling protein, MDM2, interacts with its binding partner, p53, 
through a smaller interaction interface, resembling a crevice that is lined with hydrophobic residues.
(d) The enzyme malate dehydrogenase forms a homo-dimer through a binding interface,
which is sufficiently hydrophilic that it retains waters in the bound state;
those waters are shown in red/white.
(e) The bacterial RNase barnase employs five charged residues to bind with its inhibitor barstar in one of the most stable complexes known; 
structural waters at the barnase-barstar binding interface are also shown.
(f) Lysozyme is a soluble protein with a highly amphiphilic surface that resembles a checker pattern.
For each of these diverse proteins, the susceptibility of waters in the entire protein hydration shell to the unfavorable potential, 
$\cv$ displays a marked peak.
Moreover, the peak occurs at roughly the same potential strength, $\phi^* \approx 2~\kbt$, for every protein.}
\vspace{-0.25in}
\label{fig2}
\end{figure*}

\subsection{How the hydration shells of diverse proteins respond to unfavorable potentials}
%
Is ubiquitin unique?
If not, how general is the susceptibility of protein hydration waters to an unfavorable potential? 
To address this question, we studied six additional proteins, 
spanning a range of sizes, chemical patterns, and functional roles;
see Figure~\ref{fig2} as well as Figure~S2 in the $\supp$. 
Given the importance of hydrophobic surface moieties in situating the interfacial waters at the edge of a dewetting transition, 
we first considered other proteins with well-defined hydrophobic patches.
First, we consider the fungal protein, Hydrophobin II, which is highly surface-active, and is known to self-assemble at water-vapor interfaces~\cite{hakanpaa2006hydrophobin}.
Although Hydrophobin II is charge neutral overall, the protein surface displays 10 charged residues.
In Figure~\ref{fig2}a, we show both the hydrophobic face of Hydrophobin II, which is enriched in hydrophobic residues, 
as well as the remainder of the protein, which has been shown to be super-hydrophilic~\cite{Patel:JPCB:2014}.
As with ubiquitin, the susceptibility, $\cv$, for Hydrophobin II also displays a marked peak. 
Next, we consider the human hepatitis B viral capsid protein, which has a net charge of -6, but displays an even larger hydrophobic patch than the one on Hydrophobin II (Figure~\ref{fig2}b).
That patch drives the binding of two capsid proteins to form a dimer that further assembles into a 240-protein capsid shell~\cite{capsid_struct}.
The viral capsid protein also displays a clear peak in $\cv$. 
%
%
Does the collective dewetting seen in the above proteins stem from the 
presence of extended hydrophobic patches on their surfaces?
Although many proteins possess such patches, not all of them do;
instead, most protein surfaces display chemical patterns that are amphiphilic, 
and feature only smaller hydrophobic regions.
%
The signaling protein MDM2 contains such a modest hydrophobic groove, which is nevertheless important from a functional standpoint; 
it enables MDM2 to exercise control over cellular senescence by binding to the transactivation domain of the tumor-suppresor protein p53~\cite{mdm2_struct}.
As shown in Figure~\ref{fig2}c, MDM2 also displays a peak in susceptibility, $\cv$. 

Might the large or small, but well-defined hydrophobic patches on ubiquitin, Hydrophobin II, the capsid sub-unit, and MDM2
be responsible for rendering their hydration shell waters susceptible to unfavorable perturbation?
To address this question, we study proteins that are known for being anomalously hydrophilic or charged. 
%
Malate dehydrogenase is a large hydrophilic protein with 61 charged surface residues.
The protein dimerizes into a metabolic enzyme, and plays an important role in the citric acid cycle.
The interface through which the protein monomers bind is fairly hydrophilic, featuring 5 charged residues and several other polar residues.
In fact, regions of the binding interface are so hydrophilic that they hold on their waters even in the bound state.
In other words, the binding interface features structured waters that bridge the two interacting proteins;
such bridging waters, which are observed in the crystal structure of the malate dehydrogenase dimer, 
are shown in Figure~\ref{fig2}d~\cite{bridging,mdh_struct}.
The proteins discussed previously, in contrast, feature binding interfaces that are entirely dry.
Interestingly, even for the largely hydrophilic malate dehydrogenase protein, 
we observe a peak in susceptibility, $\cv$, in response to an unfavorable perturbation (Figure~\ref{fig2}d).

Another fairly hydrophilic protein that features a charged interaction interface is barnase, 
a bacterial RNase that interacts with its inhibitor, barstar, in one of the strongest known protein-protein interactions~\cite{barnase}.
The high-affinity sub-picomolar binding between barnase and barstar is facilitated by the formation of electrostatic contacts 
between five positively charged residues on barnase and five negatively charged residues on barstar~\cite{barnase}.
Remarkably, a clear peak in susceptibility is also observed for barnase (Figure~\ref{fig2}e).
%
Finally, we study T4 lysozyme, a bacteriophage protein that catalyzes the hydrolysis of the peptidoglycan layer of bacterial cell walls~\cite{lysozyme_struct}, and has 45 charged surface residues with an overall charge of +9.
Lysozyme does not appear to participate in interactions with proteins other than its substrates,
or to possess a clear hydrophobic patch; rather, it displays a checkered pattern of hydrophobic and hydrophilic atoms (Figure~\ref{fig2}f).
As with all of the proteins studied, the hydration shell of lysozyme also displays a marked peak in susceptibility.
Our results, obtained across proteins with a diversity of sizes, biological functions, and surface chemistries, thus suggest that susceptibility to an unfavorable potential is a general property of protein hydration waters.
%

\subsection{Characterizing protein surfaces: residues vs atoms}
\begin{figure}[htb]
\centering
\includegraphics[width=0.45\textwidth]{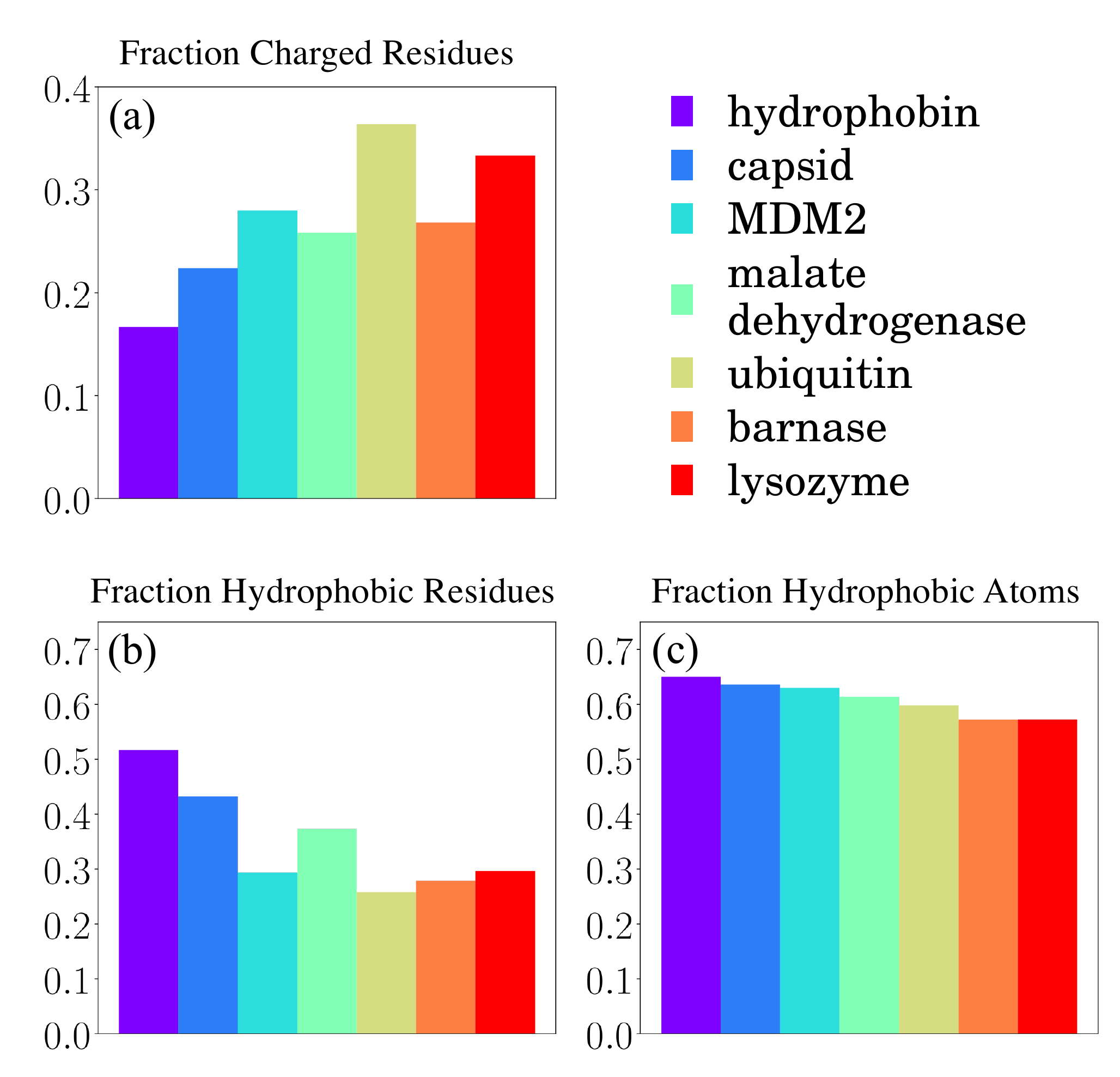} 
\caption{
An atom-centric view of diverse protein surfaces reveals that they are more hydrophobic 
than anticipated by the corresponding residue-centric view.
(a) Fraction of protein surface residues that are charged. 
%
(b) Fraction of surface residues that are hydrophobic. 
%
(c) Fraction of protein surface atoms that are hydrophobic according to the Kapcha-Rossky classification~\cite{rossky}.
}
\label{fig3}
\end{figure}

Although collective dewetting in response to an unfavorable potential may be expected for proteins that are fairly hydrophobic,
it is somewhat surprising that even the more hydrophilic and charged proteins display such behavior.
To better characterize the similarities and differences in the surface chemistries of the seven proteins discussed above, 
we plot the fraction of their surface residues that are charged and hydrophobic in Figures~\ref{fig3}a and~\ref{fig3}b, respectively. 
As expected, the more hydrophobic proteins have a smaller fraction of charged residues and a larger fraction of hydrophobic residues,
with the charge fraction ranging from 0.15 to 0.35, and the hydrophobic fraction varying from 0.5 to 0.2.
To interrogate whether the surface chemistries of these seven proteins are representative of the larger class of folded, globular proteins,
we additionally estimated these quantities for an expanded set containing a total of 20 proteins.
The results are included in Figure~S4 of the $\supp$, and highlight that the characteristics of proteins studied here are indeed representative of typical proteins.
How then do we understand the sensitivity of such diverse protein hydration shells to unfavorable potentials,
and their resemblance to extended hydrophobic surfaces?

To answer this question, we draw inspiration from work by Kapcha and Rossky, who highlighted that amino acid residues are not monolithic, 
but are instead heterogeneous, and are composed of both hydrophobic and hydrophilic atoms~\cite{rossky}.
Kapcha and Rossky thus advocate adopting an atom-centric, rather than a residue-centric view of the protein surface.
They further suggest that an atom be classified as hydrophobic only if the magnitude of its partial charge is less than 0.25 in the OPLS force field, and hydrophilic otherwise. 
Following these authors, we plot the fraction of surface atoms (not residues) that are hydrophobic in Figure~\ref{fig3}c.
Interestingly, we find that the fraction of surface atoms that are hydrophobic is not only larger than the corresponding fraction of surface residues,
but that roughly half the protein surface (or more) consists of hydrophobic atoms.
Importantly, the fraction of surface atoms that are hydrophobic is uniformly high for all proteins studied here.
To better understand these results, 
we analyzed the atomic composition of the surface residues; 
as shown in Figure~S5 of the $\supp$, roughly 80\% of atoms belonging to hydrophobic residues are hydrophobic, 
but nearly 50\% of atoms belonging to hydrophilic (polar or charged) residues are also hydrophobic.
Thus, although polar and charged surface residues do not contribute as heavily to hydrophobic surface atoms, 
they nevertheless have substantive contributions.
We note that following Kapcha and Rossky, we classify not just the protein heavy atoms, but also hydrogen atoms as being either hydrophobic or hydrophilic~\cite{rossky}.
If the protein hydrogen atoms are excluded from the analysis, the fraction of hydrophobic surface atoms is somewhat lower, 
but remains close to 0.5 as shown in Figure~S4 of the $\supp$.
Our results thus suggest that for a wide variety of proteins, roughly half the surface consists of hydrophobic atoms;
these atoms situate the protein hydration waters at the edge of a dewetting transition, making them particularly susceptible to unfavorable potentials.

\begin{figure}[htb]
\centering
\includegraphics[width=0.5\textwidth]{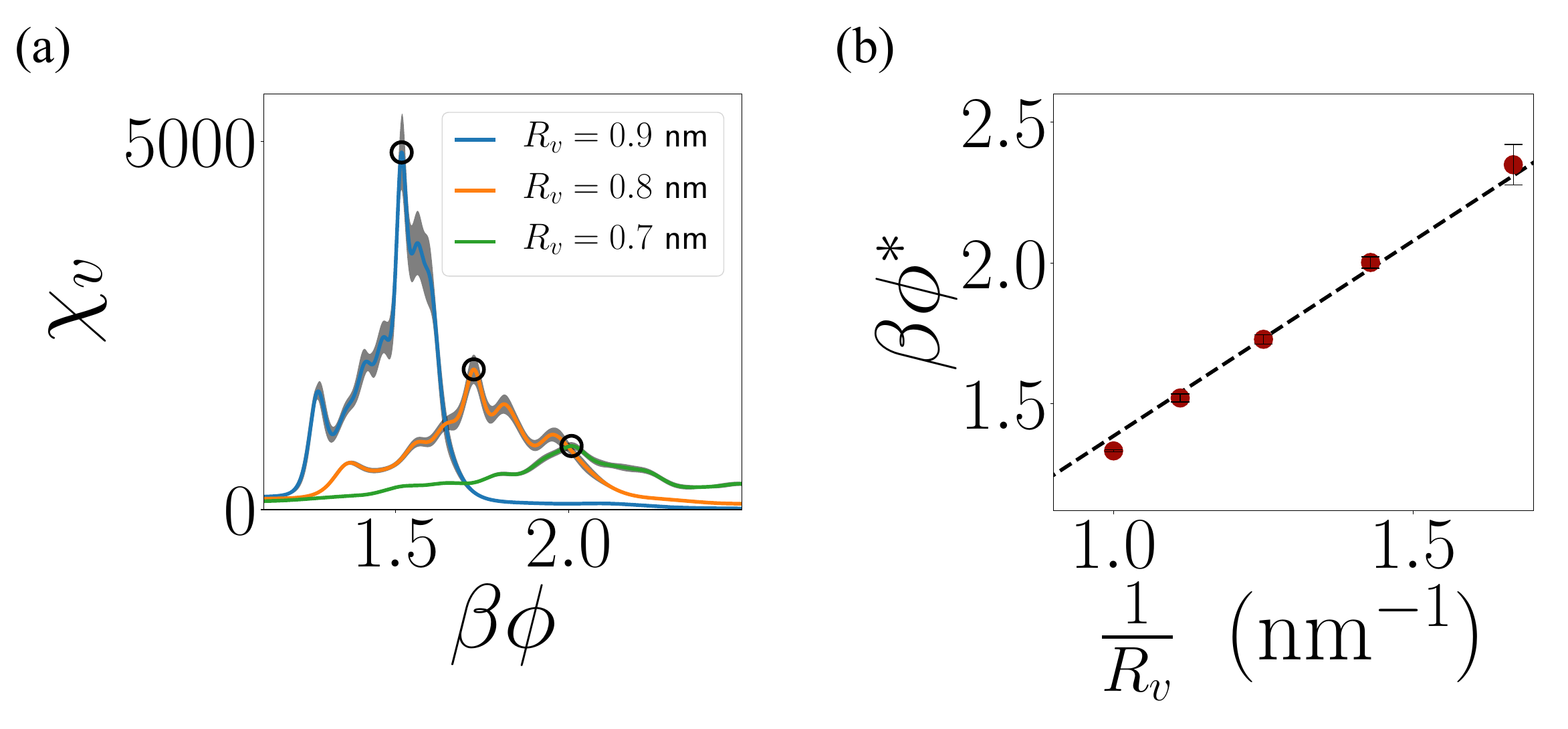} 
\caption{
How the location of the peak in susceptibility depends on the width of the protein hydration shell, $v$.  
(a) Susceptibility of the ubiquitin hydration waters to the biasing potential strength, $\phi$, 
is shown for hydration shells with different widths; 
as $R_v$ is increased, the peak in $\cv$ shifts to lower $\phi$-values. 
(b) The peak location, $\beta\phi^*$ as a function of $R_v^{-1}$ (symbols), and a linear fit to the data (line) are shown.
%
}
\label{fig4}
\end{figure}

\subsection{Strength of unfavorable potential needed to trigger dewetting}
%
Interestingly, not only do all of proteins studied here show a peak in $\cv$, 
the location of the characteristic peak in susceptibility, $\phi^*$,
is observed to be roughly $2~\kbt$ in all cases (Figure \ref{fig2}). 
To understand this observation, we first recognize that for $\phi>0$, the biasing potential, $\phi N_v$, favors configurations with lower $N_v$-values and thereby lower densities.
As a result, the biasing potential effectively lowers the pressure in the protein hydration shell to:
$P_{\rm eff} \approx P - \phi \rho_{\rm w}$, where $P$ is the system pressure and $\rho_{\rm w}$ is the molar density of liquid water.
For a sufficiently large $\phi$, the tension (negative pressure) exerted on the protein hydration waters prompts the nucleation of vapor in certain regions of $v$.
The biasing potential thus stabilizes a water-vapor interface; 
the pressure drop across this interface is related 
to the corresponding interfacial tension, $\gamma$, and the mean interfacial curvature, $\bar{\kappa}$, 
according to the Young-Laplace equation, $P - P_{\rm eff} = \gamma \bar{\kappa}$.
Moreover, because the biasing potential is only experienced by waters in $v$,
the interfacial curvature is determined by the radius, $R_v$, of the spherical sub-volumes defining $v$,
such that $\bar{\kappa} \propto 1/R_v$.
Thus, the biasing potential strength needed to trigger vapor nucleation ought to be: $\phi^* \approx (P - P_{\rm eff}) / \rho_{\rm w} \propto (\gamma / \rho_{\rm w}) (1/R_v)$.
By systematically varying $R_v$, and repeating our calculations for ubiquitin, 
we find that as $R_v$ is increased, the peak in susceptibility indeed shifts to lower values (Figure~\ref{fig4}a).
Furthermore, as shown in Figure~\ref{fig4}b, $\phi^*$ also varies linearly with $1/R_v$ as predicted; 
the best fit line through the origin has a slope of 1.38~nm, 
which is comparable to $\beta \gamma / \rho_{\rm w} = 0.45$~nm,
estimated using $\rho_{\rm w}=33$~nm$^{-3}$ and $\gamma=60.2$~mJ/nm$^2$ for SPC/E water.
%

\section{Conclusions and Outlook}
%
Proteins employ intricate topographical and chemical patterns, 
which have evolved to facilitate their many biological functions.
Although the space of such patterns is immense, 
it is likely constrained by common characteristics that 
all proteins must possess in order to function properly.
For example, all proteins must have favorable interactions with water to be soluble, 
which requires the presence of hydrophilic groups on their surfaces.
Conversely, protein surfaces must also feature hydrophobic regions that interact poorly with water, 
and provide a driving force for proteins to interact with other molecules.
In this article, we shed light on how proteins accomplish these competing goals
by balancing their overall interactions with their hydration waters.
We show that roughly half the atoms on the protein surface are hydrophobic -- a fact that can be obfuscated by focusing on surface residues rather than surface atoms.
We also find that the hydration shells of diverse proteins 
-- even those with highly charged surfaces and amphiphilic interaction interfaces -- 
are highly susceptible to an unfavorable potential.
Our results thus suggest that hydrophobic atoms on the protein surface situate its hydration waters at the edge of a dewetting transition, 
which can be triggered by an unfavorable perturbation.
Consistent with our results, signatures of collective transitions have also been observed in studies of partially hydrated proteins.
For example, Cui {\it et al.} found that partially hydrated proteins undergo a percolation transition at a critical value of protein hydration~\cite{Cui:Protein:2014}.
Similarly, in studying the uptake of water from the vapor phase by proteins, 
Debenedetti and co-workers found protein hydration to display hysteresis 
-- a hallmark of collective transitions --
between the adsorption and desorption branches of the isotherm~\cite{Palmer:JPCL:2012,Kim_2015}.
These authors also found that polar and charged residues contributed to the collective wetting of a dry protein (and the associated hysteresis)~\cite{kim2017microscopic};
correspondingly, here we find that non-polar regions of the protein give rise to the collective dewetting of a hydrated protein. 

%
We also find that the biasing potential strength needed to trigger dewetting is inversely proportional to the width of the hydration shell, 
but does not depend meaningfully on the particular protein being perturbed.
Thus, our results not only suggest that susceptibility to an unfavorable perturbation is a general feature of the hydration shells of proteins,
but also highlight that the strength of the perturbation needed to trigger dewetting is remarkably similar across different proteins.
This finding suggests a near-universal calibration of the perturbation strength across diverse proteins, i.e., by considering $\phi$ relative to $\phi^*$.
It also establishes a framework for systematically classifying how favorable the interactions between water and different parts of the protein surface are.
In particular, we expect that locations on the protein surface that dewet at low $\phi$-values (relative to $\phi^*$) 
will correspond to the most hydrophobic regions on the protein surface, 
whereas regions that retain their waters even at high $\phi$-values will be the most hydrophilic. 
Given the importance of hydrophobicity in driving protein interactions, 
regions of the protein surface that dewet most readily may 
correspond closely with patches on the protein that participate in interactions~\cite{Bogan:JMB:1998,keskin2005hot,White:2008aa,Abrams:2013:PSFB}.
We are investigating whether such hydrophobic protein regions could serve as predictors of protein interaction sites,
and plan to report our findings in a future study.
Similarly, identification of the most hydrophilic regions of the protein could facilitate the discovery of novel super-hydrophilic chemical patterns;
an understanding of what enables such patterns to have strong interactions with water could also serve as the basis for the rational design of protein non-fouling surfaces or surfaces that display super-oleophobicity underwater~\cite{nonfouling,superoleo}.

\begin{acknowledgement}
A.J.P. gratefully acknowledges financial support from the National Science Foundation (CBET 1652646, CHE 1665339, and UPENN MRSEC DMR 11-20901), and a fellowship from the Alfred P. Sloan Research Foundation.
N.B.R. was supported by the National Science Foundation grant CBET 1652646.
\end{acknowledgement}

\begin{suppinfo}
In the $\supp$, we include details of our simulations, 
information on enhanced sampling techniques that we use, 
plots supporting Figure~\ref{fig2}, and
additional analysis pertaining to the diversity of the proteins studied here. 
\end{suppinfo}


\providecommand{\latin}[1]{#1}
\makeatletter
\providecommand{\doi}
  {\begingroup\let\do\@makeother\dospecials
  \catcode`\{=1 \catcode`\}=2 \doi@aux}
\providecommand{\doi@aux}[1]{\endgroup\texttt{#1}}
\makeatother
\providecommand*\mcitethebibliography{\thebibliography}
\csname @ifundefined\endcsname{endmcitethebibliography}
  {\let\endmcitethebibliography\endthebibliography}{}

\end{document}


\section{Simulation Details}
All systems were simulated using version 4.5.3 of the GROMACS molecular dynamics package~\cite{gromacs}.
The leapfrog integrator~\cite{leapfrog} was used to integrate the equations of motion with a time-step of 2 fs. 
The oxygen-hydrogen bonds in water were constrained using the SETTLE algorithm~\cite{settle},
and all other bonds to hydrogens were constrained using the LINCS algorithm~\cite{lincs}. 
The SPC/E water model~\cite{spce} was used throughout,
and the proteins were simulated using the AMBER99SB force field~\cite{amber99sb}. 
Short-range van der Waals and Coulombic interactions were trunctated using a cut-off of 1.0 nm, and long-range electrostatics were calculated using the particle-mesh Ewald (PME) algorithm~\cite{PME}.
All simulations were performed in the canonical ensemble; the temperature was maintained at $T=300 \; K$ using the stochastic velocity-rescale thermostat~\cite{vrescale}. 
To ensure that water density fluctuations in the observation volume were not suppressed by our use of the canonical ensemble, we created a buffering water-vapor interface at the edge of the simulation box, as described elsewhere~\cite{Miller:PNAS:2007,Patel:PNAS:2011}.
The presence of a buffering interface effectively maintained the system to remain at its coexistence pressure.

\subsection{Applying Unfavorable Biasing Potentials Using INDUS}
\begin{figure*}[thb]
\centering
\includegraphics[width=0.8\textwidth]{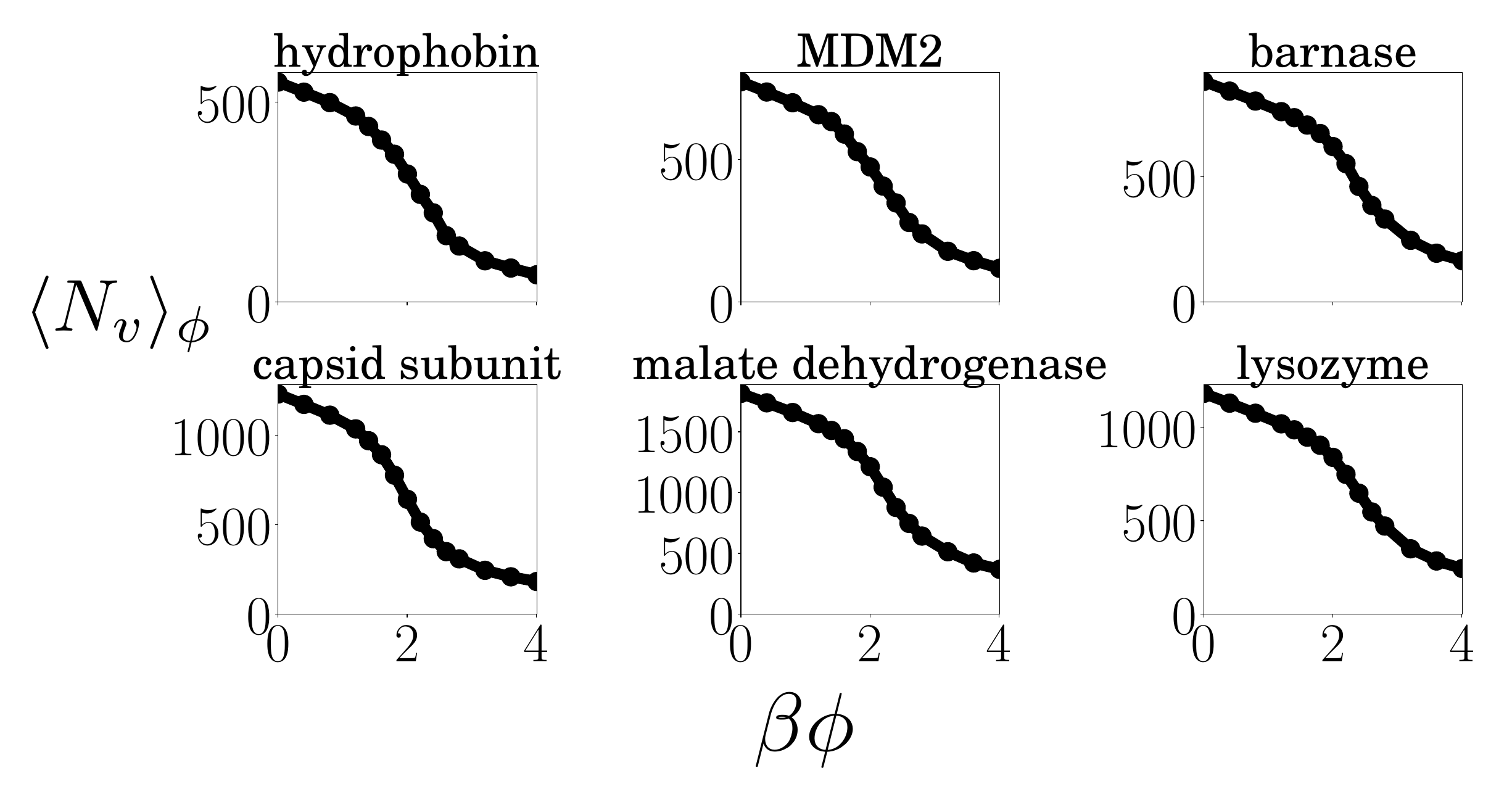} 
\caption{
In Figure~2 of the main text, the dependence of the susceptibility, $\chi_v$, on the biasing potential strength, $\phi$, is shown for six proteins.
Here, the dependence of the average coarse grained number of waters, $\langle N_v \rangle_\phi$, on $\phi$ is shown.
In each case, $\langle N_v \rangle_\phi$ displays a sigmoidal dependence on the biasing potential strength, $\phi$. 
}
\label{figS_n_v_phi}
\end{figure*}

The GROMACS package was suitably modified to bias the coarse-grained water number, $N_v$, in observation volumes of interest using the Indirect Umbrella Sampling (INDUS) prescription~\cite{Patel:JSP:2011,Patel:JPCB:2014}.
The Gaussian coarse-graining function used in INDUS was parameterized with a standard deviation of $\sigma=0.01 \; \rm{nm}$ and a trunctation length $r_c = 0.02 \; \rm{nm}$. 
In the main text and in Figure~\ref{figS_n_v_phi}, we report averages in biased ensembles; the statistics were governed by the Hamiltonian: $\mathcal{H}_\phi = \mathcal{H}_0 + \phi N_v$, where $\mathcal{H}_0$ is the unbiased Hamiltonian and $\phi$ represents the strength of the biasing potential~\cite{Patel:JPCB:2014,Xi:JCTC:2016}.
The biased ensemble averages $\langle N_v \rangle_\phi$ and $\chi_v \equiv -\partial \langle N_v \rangle_\phi / \partial (\beta \phi) = \langle \delta N_v^2 \rangle_\phi$ were obtained either by sampling directly from biased ensembles with different $\phi$-values or by reweighting the underlying unbiased free energy landscape, $F_v(N)$.
In the latter case, $F_v(N)$ was obtained using umbrella sampling by performing a series of overlapping biased simulations that were analyzed using the Multistate Bennett Acceptance Ratio (MBAR)~\cite{wham,mbar} method.
All biased simulations were run for a total of $3$~ns; the first 500 ps were discarded for equilibration. 

\subsection{SAM Surfaces}
The parameterization and setup of the SAM surfaces has been described in detail elsewhere~\cite{Godawat:PNAS:2009,Sarupria:PRL:2009,Shenogina_PRL_2009}.
Briefly, both the hydrophilic and hydrophobic SAMs were composed of 10-carbon alkyl chains that were terminated by sulfurs on one end and either an OH- or CH$_3$- head group on the other. 
Each SAM surface had a total of 224 chains; the sulfur atoms were arranged on a hexagonal lattice with a spacing of 0.5 nm between adjacent chains. 
Both SAM surfaces had a cross-sectional area of 7 x 7 nm in the $y$ and $z$ dimensions and a height of approximately 1.3 nm in the $x$ dimension. 
The head groups of the SAM surfaces were solvated with a total of 6,464 water molecules; the corresponding water slab had a width of approximately 4 nm. 
The simulation boxes were then extended in the $x$ direction to create a buffering liquid-vapor interface.
%
The solvated SAM surfaces were equilibrated for $1$~ns to produce the starting structures for the biased simulations.
A cylindrical observation volume, $v$, with a height of $0.3$ nm in the $x$ dimension and a radius of $2.0$ nm in the $y$-$z$ plane
was placed at the SAM-water interface. 
The coarse-grained number of water molecules, $N_v$ in $v$, was biased using a harmonic potential with a spring constant of $\kappa = 0.24$ and $0.34$ $\frac{kJ}{mol}$ for the CH$_3$- and OH- terminated SAMs, respectively. 
For each SAM, a total of 22 biased simulations were employed to sample the entire range of $N_v$ and to estimate $F_v(N)$.
%

\subsection{Protein Systems}
The following crystal structures were used to prepare the seven proteins described in the main text: Ubiquitin (PDB: 1UBQ) \cite{ubiq_struct}, barnase (PDB: 1BRS) \cite{barnase_struct}, hepatitis B viral capsid (PDB: 1QGT) \cite{capsid_struct}, MDM2 (PDB: 1YCR) \cite{mdm2_struct}, bacteriophage T4 lysozyme (PDB: 253L) \cite{lysozyme_struct}, hydrophobin II (PDB: 2B97) \cite{hfb_struct}, and malate dehydrogenase (PDB: 3HHP) \cite{mdh_struct}. 
All protein systems were prepared from their PDB structures by first removing crystallographic waters and co-solutes. 
Each system was then placed in a cubic box; the side lengths were chosen so that all protein atoms were at least 1.6 nm from the box edges. 
The proteins were solvated using the GROMACS `genbox' utility. Sodium or chloride counter ions were added, if needed, to maintain a net neutral charge. 
The simulation boxes were extended in the $z$ direction to create a buffering liquid-vapor interface.
Water slabs were maintained at the centers of the simulation boxes by restraining the $z$ component of their centers of mass to a dummy atom in the center of the box using a harmonic restraint with a spring constant of $1000$~kJ/mol/nm$^2$. 
The initial configurations for all protein systems were prepared using steepest-descent energy minimization followed by 1~ns of equilibration. 
These configurations were used as the starting structures for the biased simulations. 
The observation volume, $v$, was defined as the union of spherical sub-volumes centered on the initial positions of the protein heavy (non-hydrogen) atoms. 
Each spherical sub-volume was chosen to have the same radius, $R_v$. $R_v$ was chosen to be $0.6$ nm for all proteins. For ubiquitin, $R_v$ was varied systematically from $0.4$ nm to $1.0$ nm.
To ensure that the protein hydration shell continued to overlap with $v$ throughout the biased simulations,
the protein heavy atoms were position-restrained to their initial positions by applying a harmonic potential with a spring constant of $\kappa=1000$~kJ/mol/nm$^2$. 
%

\section{Response of Protein Hydration Waters to an Unfavorable Potential}

\section{Properties of the Protein Surfaces Studied}
\begin{figure*}[thb]
\centering
\includegraphics[width=0.7\textwidth]{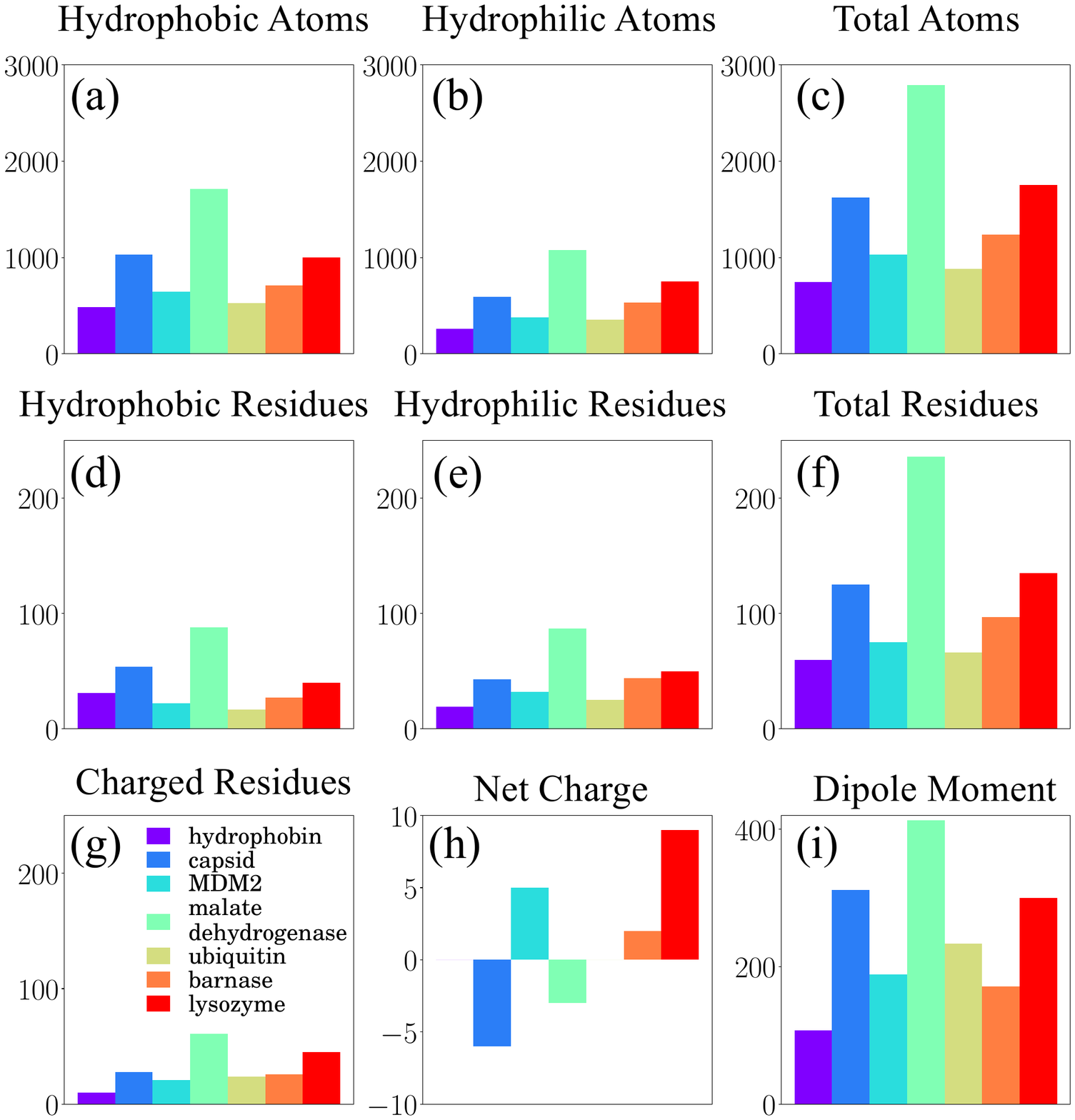} 
\vspace{-0.6in}
\caption{
The seven proteins studied here display a range of surface chemistries. 
The number of hydrophobic, hydrophilic, and total surface atoms for each protein are shown in panels (a), (b), and (c), respectively;
the corresponding number of surface residues are shown in panels (d), (e), and (f).
The total number of charged surface residues (positive or negative), the net charge, and the total dipole moment (in Debye units) of the proteins are shown in panels (g), (h), and (i), respectively.
}
\label{figS1}
\end{figure*}

%
Figure~\ref{figS1} highlights the differences in the surface chemistries of the seven protein systems described in the main text.
These proteins were chosen to span the broad range of surface chemistries representative of globular proteins. 
\vspace{-0.6in}

\subsection{Determining Protein Surface Atoms, Heavy Atoms and Residues}
We identified protein surface atoms by their exposure to solvent. 
To determine solvent exposure, the average number of water oxygens within 0.6 nm of each protein heavy atom was calculated from a 2.5 ns equilibrium simulation; those with 6 water oxygens or more were classified as surface heavy atoms. 
Hydrogen atoms bonded to any surface heavy atoms were also considered to be surface atoms.
Moreover, any amino acid residue containing at least one surface atom was categorized as a surface residue.
Although many other reasonable ways for making the above classifications exist, 
we do not believe that they are likely influence our qualitative findings.
The total surface atoms and surface residues for the seven proteins 
studied in the main text are shown in Figures~\ref{figS1}c and~\ref{figS1}f, respectively.

\begin{figure*}[thb]
\centering
\includegraphics[width=0.8\textwidth]{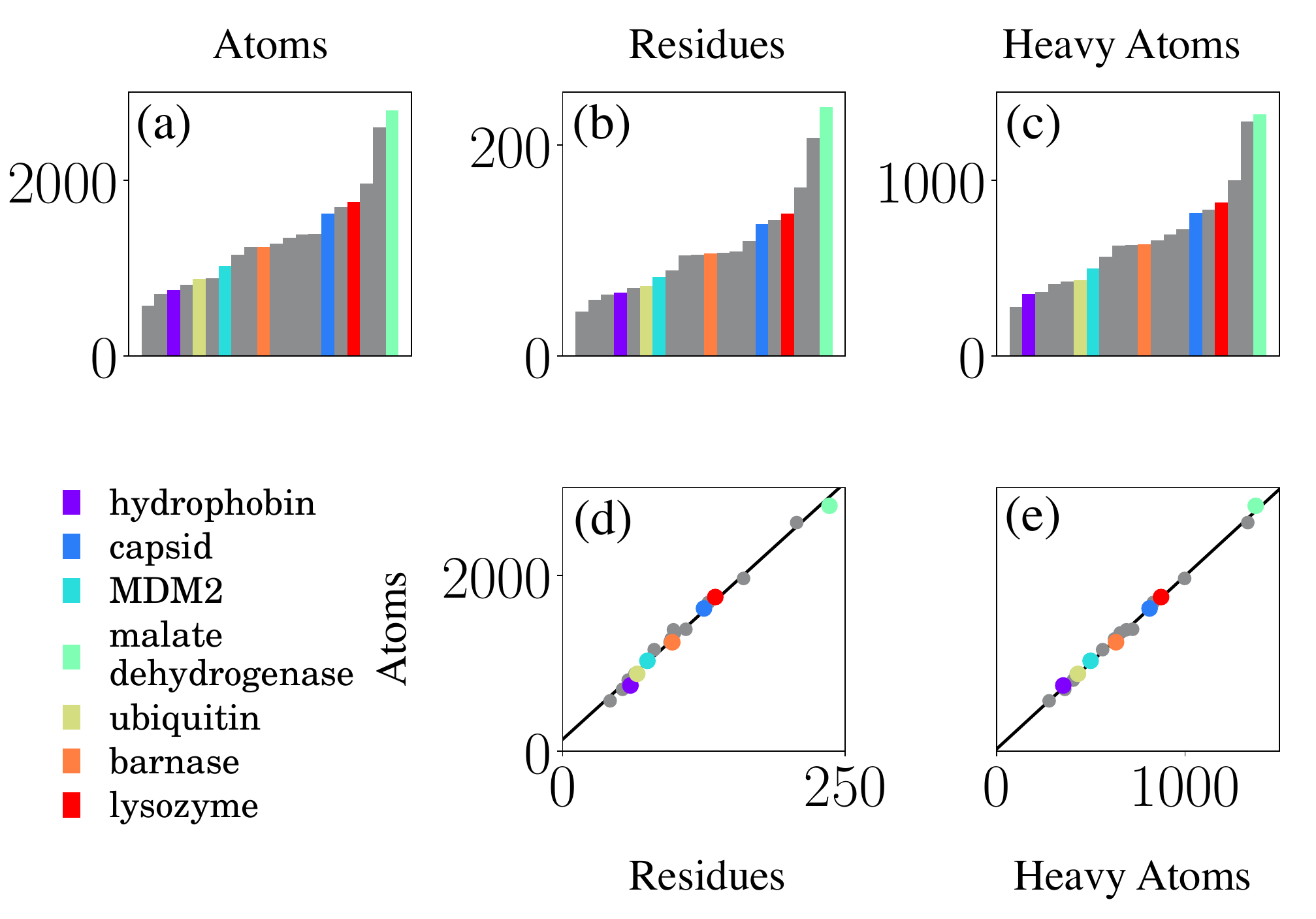} 
\caption{
The seven proteins studied span a wide range of sizes that is representative of globular proteins.
Results for the studied systems are shown in color, whereas those for 13 additional proteins are shown in gray, 
and are included for comparison. 
The number of protein surface atoms, surface residues, and surface heavy atoms (non-hydrogen atoms) are shown in panels (a), (b), and (c), respectively.
Panels (d) and (e) illustrate that the number of surface atoms vary linearly with the number of surface residues and the number of surface heavy atoms, respectively, highlighting that all three quantities contain the same information about the relative sizes of the proteins.
}
\label{figS2}
\end{figure*}

\subsection{Chemical Classification of Protein Surface Atoms and Residues}
The protein surface atoms were classified as either hydrophobic or hydrophilic according to the atom-wise hydropathy scale developed by Kapcha and Rossky~\cite{rossky}. 
The number of protein surface atoms that are hydrophobic and hydrophilic are shown in Figures~\ref{figS1}a and~\ref{figS1}b, respectively, for the seven proteins studied in the main text.
Amongst the surface residues, the following were classified as hydrophobic: alanine, valine, leucine, isoleucine, phenylalanine, 
proline
and tryptophan.
The remaining residues were classified as hydrophilic.
%
Out of the hydrophilic residues, those with a net charge of -1 or +1 were classified as charged; these residues include aspartate, glutamate, arginine, lysine, and protonated histidine.
A pH of 7 was assumed, and protonation states were assigned by the GROMACS utility `pdb2gmx';
protonation states of histidine residues were assigned according to their positions in the crystal structure. 
The total number of hydrophobic, hydrophilic, and charged residues for the seven proteins 
studied in the main text are shown in Figures~\ref{figS1}d,~\ref{figS1}e, and~\ref{figS1}g, respectively.
The net charge on the protein surface was determined by summing the charges of all surface residues that are charged, and is shown in Figures~\ref{figS1}h.
Protein dipole moments were calculated from the crystal structures using the Protein Dipole Moments server (http://bip.weizmann.ac.il/dipol)~\cite{dipole_server}, and are shown in Debye units in Figure \ref{figS1}i.
%

\subsection{Comparison with an Expanded Protein Library}
To determine whether the range of surface chemistries for the seven proteins studied here is characteristic of globular proteins, we examined the surface properties of an expanded library containing 13 additional proteins. 
These proteins all participate in protein-protein intermolecular interactions; 
otherwise, they were chosen without regard to their surface chemistries
and should represent the typical diversity of protein surface chemistries. 
The PDB ID's of the additional proteins are: 1AUO \cite{xtal_1auo}, 1CMB \cite{xtal_1cmb}, 1GVP \cite{xtal_1gvp}, 1HJR \cite{xtal_1hjr}, 1MSB \cite{xtal_1msb}, 1PP2 \cite{xtal_1pp2}, 1UTG \cite{xtal_1utg}, 1WR1 \cite{xtal_1wr1}, 2K6D \cite{xtal_2k6d}, 2QHO \cite{xtal_2qho}, 2RSP \cite{xtal_2rsp}, 2TSC \cite{xtal_2tsc}, and 2Z59 \cite{xtal_2z59}.
These systems span a range of sizes; the number of surface atoms, surface residues, and surface heavy atoms for proteins in this expanded library are shown in Figures~\ref{figS2}a-c. 
Trends in protein sizes were also observed to be consistent across the three metrics; see Figures~\ref{figS2}d,e.

\begin{figure*}[thb]
\centering
\includegraphics[width=0.8\textwidth]{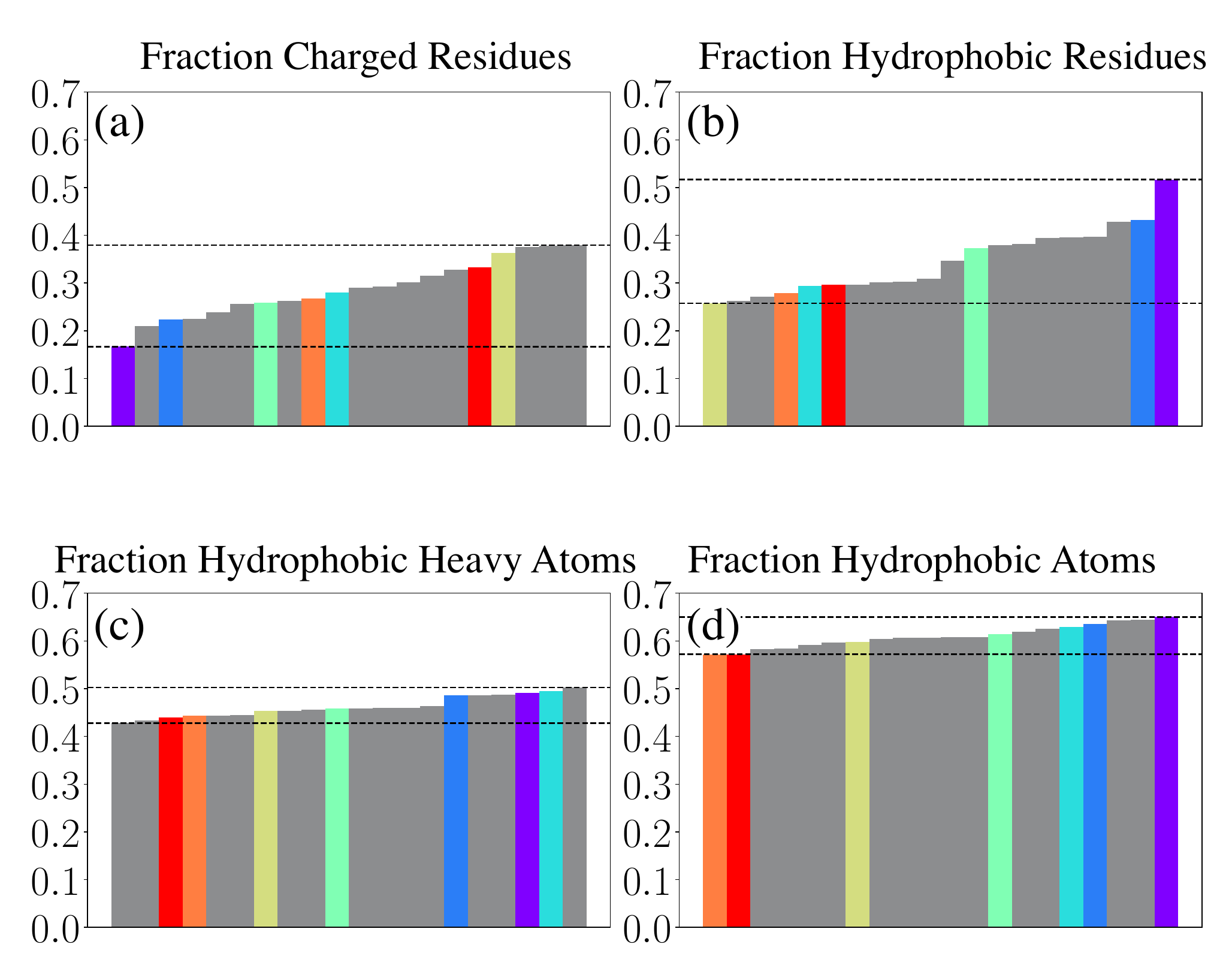} 
\caption{
The hydrophobicity of a protein surface depends on whether entire residues or individual atoms are considered. 
The fraction of surface residues that are charged (a) or hydrophobic (b) varies substantially across different proteins; roughly 15 to 35\% of the surface residues are charged, and 25 to 50\% of the surface residues are hydrophobic. 
The fraction of surface heavy atoms that are hydrophobic (c), or the fraction of all surface atoms that are hydrophobic (d)
are not only higher than the fraction of surface residues that are hydrophobic, they are remarkably similar across all the proteins;
the hydrophobic fraction of surface heavy atoms spans ranges from roughly 45 to 50\%, 
and that of the all surface atoms spans from 55 to 65\%.
The color scheme is the same as in Figure~\ref{figS2}.
}
\label{figS3}
\end{figure*}

\subsection{Residue- vs Atom-centric Characterization of Protein Surfaces}
In Figure 3 of the main text, we showed that the fraction of hydrophobic surface atoms is larger than the fraction of hydrophobic surface residues. Additionally, the fraction of per-atom hydrophobic composition is remarkably similar across the proteins studied, while the per-residue composition is not.
As shown in Figure~\ref{figS3}, these trends are also observed in the expanded library of proteins.
Figures~\ref{figS3}a and~\ref{figS3}b show the fraction of surface residues that are either charged or hydrophobic, respectively;
both fractions vary by nearly 20 - 25\% across the proteins considered.
However, the variation in the fraction of surface atoms (or heavy atoms) that are hydrophobic is only about 5 - 10\% (Figures~\ref{figS3}c,d).
Moreover, the fraction surface atoms that are hydrophobic is greater than 50\% for all protein systems considered, and is consistently greater than the fraction of surface residues that are hydrophobic.

\begin{figure*}[thb]
\centering
\includegraphics[width=0.7\textwidth]{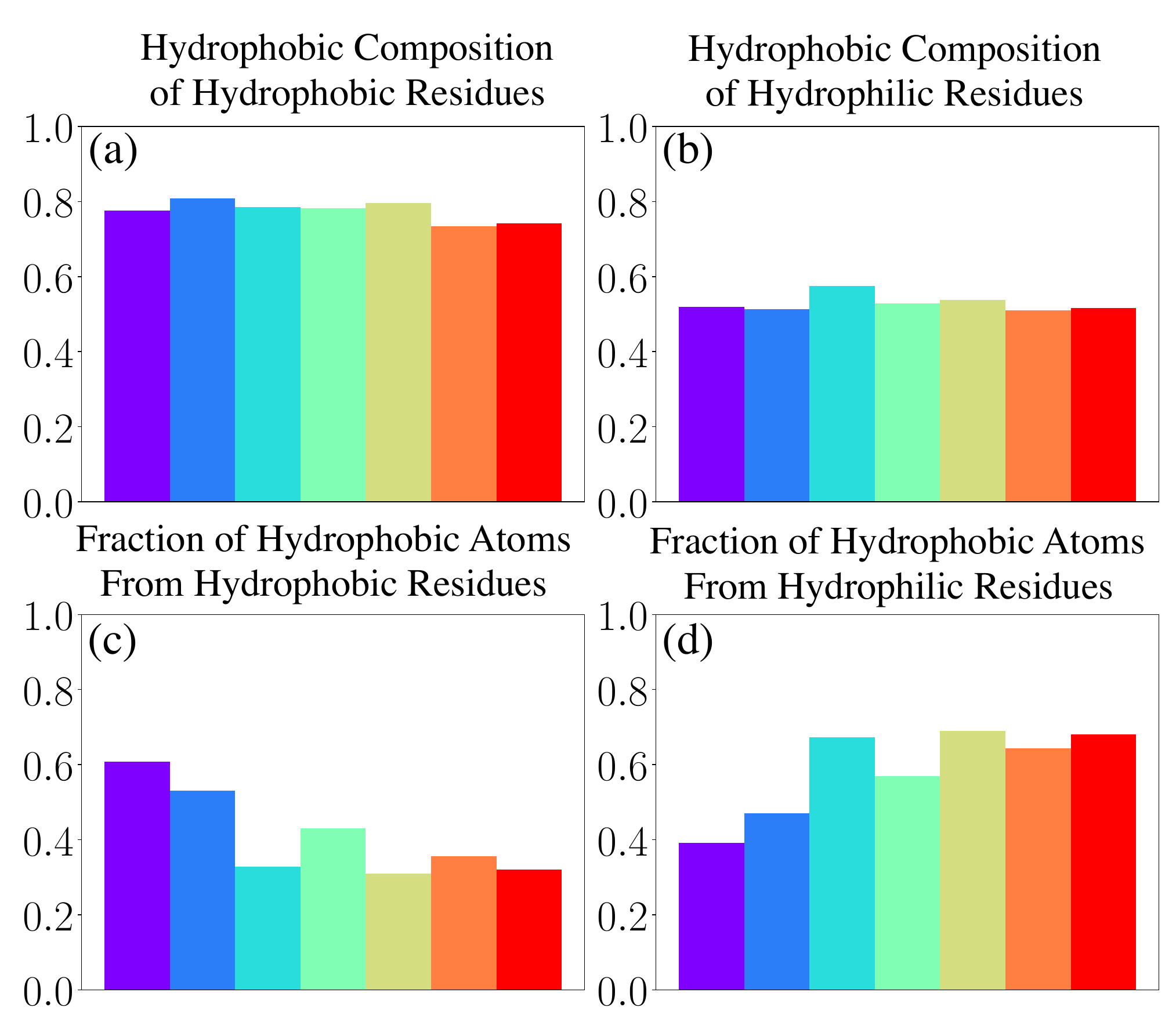} 
\caption{
The differences in the hydrophobic fraction of surface residues and that of surface atoms can be understood by considering the atomic composition of the surface residues. 
The fraction of atoms from either hydrophobic (a) or hydrophilic (b) surface residues that are hydrophobic. 
As expected, hydrophobic residues are composed primarily of hydrophobic atoms; roughly 80\% of their atoms are hydrophobic.
However, nearly 50\% of the atoms belonging to hydrophilic surface residues are also hydrophobic.
Thus, hydrophilic residues contribute substantially to the overall hydrophobicity of the protein surface. 
Consequently, as we consider proteins with surfaces that have a decreasing fraction of hydrophobic residues and an increasing fraction hydrophilic (and charged) residues,
hydrophobic residues contribute less (c) and hydrophilic residues contribute more (d) to the overall fraction of hydrophobic surface atoms.
The color scheme is the same as in Figure~\ref{figS2}.
}
\label{figS4}
\end{figure*}

To better understand the discrepency between the fraction of hydrophobic surface atoms and hydrophobic surface residues,
we considered into the atomic composition of the surface residues. 
Figure \ref{figS4}a shows that nearly 80\% of the surface atoms belonging to hydrophobic residues are themselves hydrophobic; 
hydrophobic residues are composed mostly of hydrophobic atoms (the remaining 20\% are primarily backbone atoms).
Surprisingly, nearly 50\% of the surface atoms that belong to hydrophilic residues are also hydrophobic, as shown in Figure \ref{figS4}b. 
This observation helps explains why seemingly hydrophilic proteins 
with a large fraction of hydrophilic and/or charged surface residues 
contain a large fraction of surface hydrophobic atoms. 
%
The fractions of the hydrophobic surface atoms that belong to either hydrophobic or hydrophilic surface residues are shown in Figures~\ref{figS4}c,d, and highlight that 
as proteins become seemingly more hydrophilic (i.e., have a larger fraction of surface residues that are hydrophilic),
they contribute an increasingly larger fraction of hydrophobic atoms to the protein surface.

\providecommand{\latin}[1]{#1}
\makeatletter
\providecommand{\doi}
  {\begingroup\let\do\@makeother\dospecials
  \catcode`\{=1 \catcode`\}=2 \doi@aux}
\providecommand{\doi@aux}[1]{\endgroup\texttt{#1}}
\makeatother
\providecommand*\mcitethebibliography{\thebibliography}
\csname @ifundefined\endcsname{endmcitethebibliography}
  {\let\endmcitethebibliography\endthebibliography}{}